\documentclass[prl,twocolumn,superscriptaddress,showpacs]{revtex4}
\usepackage{amsmath}
\usepackage{amsfonts}
\usepackage{graphicx}
\usepackage{pst-all}
\usepackage{psfrag}

\begin{document}

\title{Competing Topological Orders in the $\nu=12/5$ Quantum Hall State}
\author{Parsa Bonderson}
\affiliation{Station Q, Microsoft Research, Santa Barbara, California 93106-6105, USA}
\author{Adrian E. Feiguin}
\affiliation{Station Q, Microsoft Research, Santa Barbara, California 93106-6105, USA}
\affiliation{Condensed Matter Theory Center, University of Maryland, College Park, Maryland 20742, USA}
\affiliation{Department of Physics and Astronomy, University of Wyoming, Laramie, Wyoming 82071, USA}
\author{Gunnar M\"oller}
\affiliation{Theory of Condensed Matter Group, Cavendish Laboratory, J. J. Thomson Avenue, Cambridge CB3 0HE, United Kingdom}
\author{J. K. Slingerland}
\affiliation{School of Theoretical Physics, Dublin Institute for Advanced Studies, 10 Burlington Rd, Dublin 4, Ireland}
\affiliation{Department of Mathematical Physics, National University of Ireland Maynooth, Ireland}
\date{\today}

\begin{abstract}
We provide numerical evidence that a $p_{x}-i p_{y}$ paired Bonderson-Slingerland (BS) non-Abelian hierarchy state is a strong candidate for the observed $\nu=12/5$ quantum Hall plateau. We confirm the existence of a gapped incompressible $\nu = 12/5$ quantum Hall state with shift $S=2$ on the sphere, matching that of the BS state. The exact ground state of the Coulomb interaction at $S=2$ is shown to have large overlap with the BS trial wave function. Larger overlaps are obtained with BS-type wave functions that are hierarchical descendants of general $p_{x}-i p_{y}$ weakly-paired states at $\nu=5/2$. We perform a finite size scaling analysis of the ground state energies for $\nu=12/5$ states at shifts corresponding to the BS ($S=2$) and $3$-clustered Read-Rezayi ($S=-2$) universality classes. This analysis reveals very tight competition between these two non-Abelian topological orders.
\end{abstract}

\maketitle


Fractional quantum Hall (FQH) physics in the lowest Landau level (LLL) is well understood in terms of the Laughlin states~\cite{Laughlin83} and the Haldane-Halperin (HH) hierarchy states~\cite{Haldane83,Halperin84}, which can equivalently be described using Jain's composite fermion (CF) approach~\cite{Jain89,Read90}. The first appearance~\cite{Willett87,Pan99} of an even-denominator FQH plateau at $\nu=5/2$ made it clear that the physics of the second Landau level (2LL) could be even more interesting. Numerical studies~\cite{Morf98,Rezayi00,Moller08a,Feiguin08a,Storni08} support the non-Abelian $p_{x}-ip_{y}$ paired Moore-Read (MR) state~\cite{Moore91} (and its particle-hole conjugate, $\overline{\text{MR}}$) as the correct description of the $\nu=5/2$ FQH state.
At first, it seemed that this exceptional filling fraction was just an anomaly that appeared amongst other ``standard'' odd-denominator FQH states at $\nu = 7/3$, $8/3$, and $14/5$. Later, a well-developed FQH state at $\nu = 12/5$ also emerged~\cite{Xia04,Pan08,Kumar10}, and it was numerically shown that, in addition to the Abelian HH state,
the particle-hole conjugate of the non-Abelian $3$-clustered Read-Rezayi (RR) state~\cite{Read99,Rezayi09a} is also a viable candidate for $\nu = 12/5$.

Recently, a non-Abelian hierarchy of states constructed over the $\nu = 5/2$ MR state was proposed~\cite{Bonderson07d} as candidates for all observed 2LL FQH plateaus, as well as filling fractions with features of potentially developing FQH states, such as $\nu=19/8$~\cite{Xia04,Pan08,Kumar10}. These Bonderson-Slingerland (BS) states exhibit the same $p_{x}-ip_{y}$ pairing and non-Abelian nature as the parent MR state. This indicates that the pairing physics of the $\nu=5/2$ ``anomaly'' could in fact also characterize other 2LL states, via the hierarchy/CF physics seen in the LLL.

Non-Abelian FQH states represent entirely new phases of matter and could potentially be used for topologically-protected quantum computation (TQC)~\cite{Kitaev03,Freedman02a}. The $3$-clustered $\overline{\text{RR}}$ state would be ideal for TQC as it can provide computationally universal gates from braiding alone, whereas the $p_{x}-ip_{y}$ paired states (e.g. MR and BS) cannot, requiring at least one supplementary unprotected gate. Hence, the BS and $\overline{\text{RR}}$ $\nu=12/5$ states have vastly different levels of utility for quantum computation. (The Abelian HH state is essentially useless for TQC).

In this Letter, we provide numerical evidence establishing a BS state as a competitive candidate at $\nu=12/5$. To study its validity and compare it with the HH and $\overline{\text{RR}}$ candidates, we use a combination of powerful numerical techniques on the sphere: exact diagonalization, variational Monte Carlo, and the density matrix renormalization group method of~\cite{Feiguin08a} (see also~\cite{Shibata01}).

The BS state that we study as a candidate for $\nu=12/5$ can also be described using a CF type formulation~\cite{Bonderson07d}, with ground-state trial wave function~\footnote{The last equality does not hold strictly since the LLL projection is applied in different ways, but the  wave functions written represent the same universality class.}
\begin{equation}
\Psi_{\frac{2}{5}}^{\left( \text{BS} \right)} \!  = \! \mathcal{P}_{LLL} \!\! \left\{ \text{Pf} \left[ \frac{1}{z_{i} - z_{j}} \right] \chi_{1}^{3} \chi_{-2} \right\}
\! = \! \Psi_{1}^{\left( \text{MR} \right)} \Psi_{\frac{2}{3}}^{\left( \text{CF} \right)}
\label{eq:BS_12_5}
\end{equation}
where $\mathcal{P}_{LLL}$ is the projection onto the LLL, $\chi_{n}$ is the wave function of $n$ filled Landau levels ($n<0$ corresponding to negative flux), $\Psi_{1}^{\left( \text{MR} \right)}$ is the bosonic $\nu=1$ MR ground-state wave function~\cite{Moore91}, and $\Psi_{\frac{2}{3}}^{\left( \text{CF} \right)}$ is the $\nu=2/3$ CF ground-state wave function~\cite{Jain89}. This BS state has shift $S=2$ on the sphere, where $N_{\phi} = \nu^{-1} N_{e}-S$ is the relation between the number of flux quanta $N_{\phi}$ and the number of electrons $N_{e}$. The HH and $\overline{\text{RR}}$ states at $\nu=12/5$, respectively, have $S=4$ and $-2$ on the sphere.

\begin{figure}[t!]
\begin{center}
  \includegraphics[width=70mm]{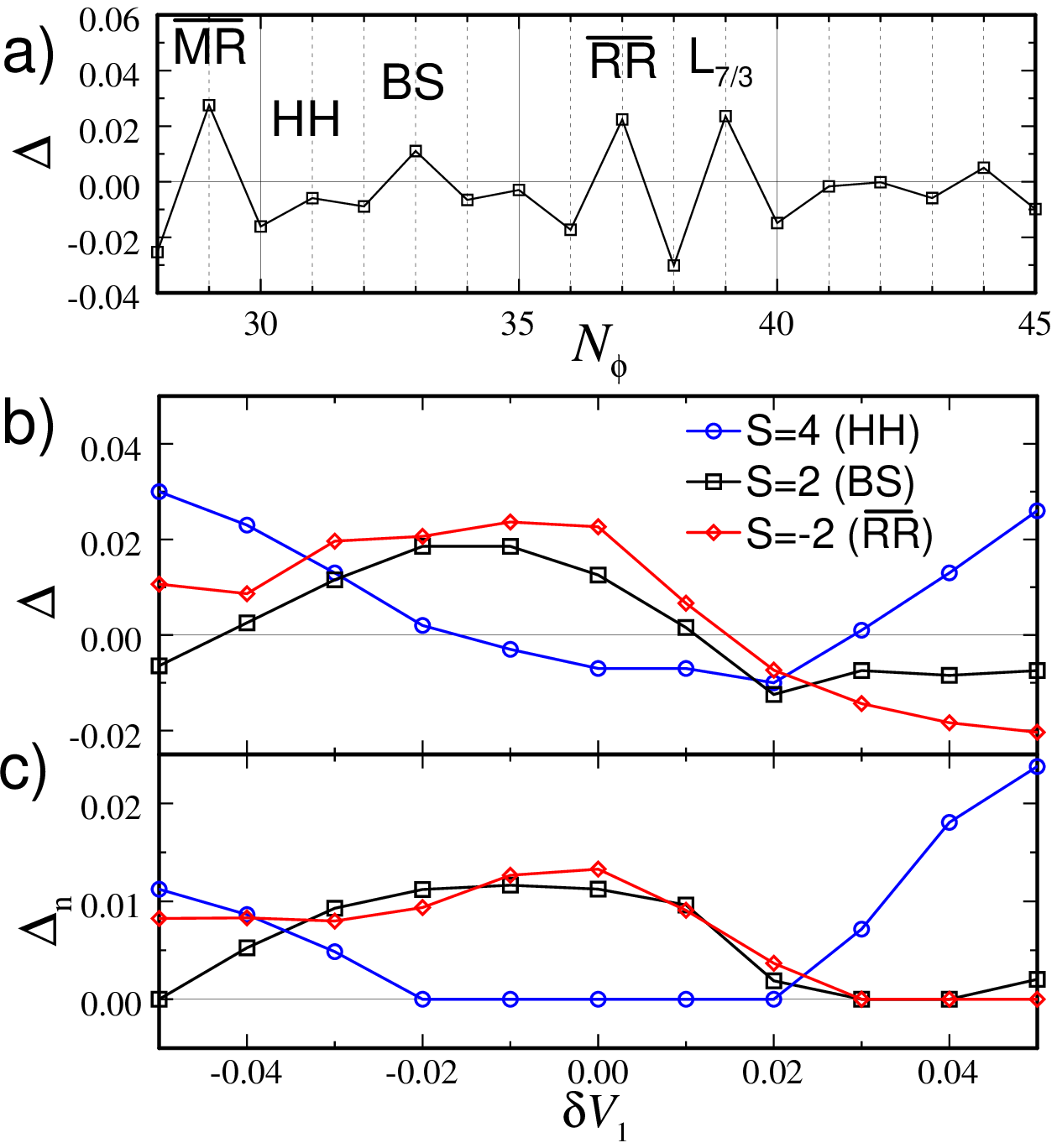}
  \caption{Excitation gaps (in units of $e^{2} / \ell_{0}$, where $\ell_0=\sqrt{\hbar c/eB}$ is the magnetic length) for systems of $N_e=14$ electrons. a) Charge gaps $\Delta$ as a function of magnetic flux $N_{\phi}$. Values corresponding to candidate states are labeled. b) and c) Charge gaps $\Delta$ and neutral gaps $\Delta_n$ under variation of the pseudopotential $V_{1}$ around the Coulomb point at fluxes corresponding to $\nu=12/5$ at shifts $S=4$, $2$, and $-2$.}
  \label{gaps}
\end{center}
\end{figure}

An important signature of a FQH state with ground-state at $N_{\phi}$ is the existence of a charge excitation gap
\begin{equation}
\Delta \left( N_{\phi}\right) = E_{N_{\phi}+1} + E_{N_{\phi}-1} - 2E_{N_{\phi}}
,
\end{equation}
where $E_{N}$ is the ground-state energy of the system with $N$ fluxes. $\Delta / n$ is the energy to create a quasihole-quasielectron pair, where $n$ fundamental quasiholes are produced per flux. ($n=2$ for the HH, BS, and $\overline{\text{RR}}$ states at $\nu=12/5$.) The existence of charge gaps may help to identify the shifts of competitive candidate states.

In Fig.~\ref{gaps}a), we show a scan of the charge gap as a function of $N_{\phi}$, for $N_e=14$ at the Coulomb point. We can identify different states according to their shifts and label the $\nu=5/2$ $\overline{\text{MR}}$ state, the $\nu=12/5$ HH, BS, and $\overline{\text{RR}}$ states, and the $\nu=7/3$ Laughlin (L$_{7/3}$) state. We find gaps for $\nu=12/5$ states at $S=2$ and $S=-2$, but not at $S=4$.
A similar scan for $N_e=10$ and $12$, which was also performed in~\cite{Wojs08}, reveals a charge gap for a $\nu=12/5$ state with $S=2$. Possible charge gaps for $\nu=12/5$ states with $S=4$ and $S=-2$ are obscured at these smaller system sizes by aliasing with the $\overline{\text{MR}}$ and L$_{7/3}$ states, respectively~\footnote{Systems on the sphere can exhibit ``aliasing'' competition between states with different filling fractions that occupy the same or adjacent values of $N_\phi$ for a given $N_e$.}.

In Fig.~\ref{gaps}b), we show the behavior of the charge gap as a function of the pseudopotential $V_1$ varied around the Coulomb point ($\delta V_1=0$), for $N_e =14$. This exhibits the generally observed behavior where increasing $V_1$ destroys the non-Abelian clustered states (BS and $\overline{\text{RR}}$) and stabilizes the Abelian state (HH). We note that the $S=2$ and $-2$ states both show a strong charge gap in the same range of $\delta V_1$, including the Coulomb point.

In Fig.~\ref{gaps}c), we show the neutral gap $\Delta_n$, i.e. the gap above the ground state at a given $N_{\phi}$, as a function of the pseudopotential $V_1$ varied around the Coulomb point, for $N_e =14$. The behavior is very similar to that of the charge gaps, but the neutral gaps remain open closer to the transition (around $\delta V_1= 0.025$) from the region where the $S=2$ and $-2$ states are stabilized and where the $S=4$ state is stabilized.

\begin{figure}[t!]
\begin{center}
\psfrag{ylabel}[c][t][1.25]{$|\langle \Psi^\text{(BS)}|\Psi_\text{exact}\rangle|^2$}
\psfrag{xlabel1}[c][b][1.25]{$\delta V_1$}
\psfrag{xlabel2}[c][b][1.25]{$\delta V_1$}
\psfrag{N=8}[l][l][0.9]{$N_e=8$}
\psfrag{N=10}[l][l][0.9]{$N_e=10$}
\psfrag{N=12}[l][l][0.9]{$N_e=12$}
\psfrag{N=14}[l][l][0.9]{$N_e=14$}
\psfrag{inset12}[l][l][0.5]{$N_e=12$}
\psfrag{inset14}[l][l][0.5]{$N_e=14$}
\psfrag{inset16}[l][l][0.5]{$N_e=16$}
  \includegraphics[width=0.95\columnwidth]{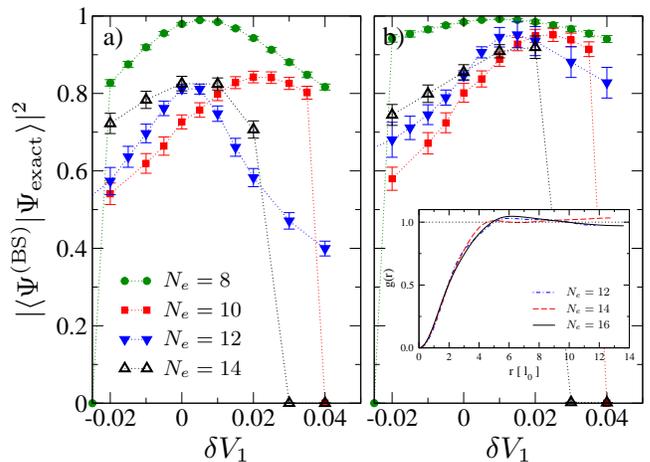}
  \caption{Squared overlaps for $N_e=8$, $10$, $12$, and $14$ between the exact diagonalization
ground-state and: a) the BS ground-state wave function of Eq.~(\ref{eq:BS_12_5}), and b) the BS ground-state with optimized pair-wave function of Eq.~(\ref{eq:generalG}). Error bars represent the statistical uncertainty of the Monte-Carlo sampling of the overlap integral. \textbf{Inset:} The pair correlation function for the $\nu=12/5$ state with $S=2$ at the Coulomb point for $N_e=12$, $14$, and $16$.}
  \label{fig:overlaps}
\end{center}
\end{figure}

To further investigate the characteristics of the $\nu=12/5$ state with $S=2$, we calculate the pair correlation function $g\left(r\right)$ obtained from exact diagonalization. The results at the Coulomb point are displayed in the inset to Fig.~\ref{fig:overlaps} and show strongly damped long-distance oscillations indicative of an incompressible state.
We also see a slight ``shoulder'' at small $r$, which becomes more pronounced as $\delta V_{1}$ decreases. Such shoulders are present for the MR and RR states~\cite{Read96,Read99}, and are considered a characteristic of non-Abelian clustered states.

To provide direct evidence that the incompressible states we find at $S=2$ (for small system sizes) are in the BS state's universality class, we compute the overlap of the trial wave function of Eq.~(\ref{eq:BS_12_5}) with the exact ground-state wave function obtained at $S=2$ on the sphere. As shown in Fig.~\ref{fig:overlaps}a), these overlaps reach as high as $0.989(2)$ for $N_{e}=8$ and remain as large as $0.83(2)$ at $\delta V_1 = 0$ for $N_{e}=14$, the largest system considered, where there are $11,463$ states in the $L^2=0$ subspace. Again, we manipulated
the first pseudopotential coefficient $V_1$ around the Coulomb potential of a thin
2DEG in 2LL to obtain a simple parametrization of the relevant interactions. The largest values of
the overlap are obtained at slightly positive values of $\delta V_1 \simeq 0.01$.
Our numerical evaluation of the overlap integrals were undertaken by Monte-Carlo sampling in position space.
The composite fermion part $\Psi^\text{(CF)}_\frac{2}{3}$ of the trial state in Eq.~(\ref{eq:BS_12_5})
was generated as a Slater determinant of individually projected CF orbitals~\cite{Jain97} at negative effective
flux~\cite{Moller05}. The rate-limiting step is the evaluation of the exact wave function, which
requires calculating a number of Slater determinants equal to the dimension $D_{L_{z}=0}$ of the $L_z=0$ subspace ($D_{L_{z}=0}\simeq 1.9\times 10^7$ for $N_{e}=14$).

The MR state may be regarded as one representative of an entire family of weakly paired CF states~\cite{Read00,Moller08a}. This is also true for the BS states that are derived from the MR state. Other representatives in either class of paired states can be obtained
explicitly by varying the pair wave function~\cite{Moller08a}. In this approach,
we introduce a number of variational parameters $g_k$ and replace the $p_x -i p_y$ pair wave function as follows:
\begin{equation}
\label{eq:generalG}
\text{Pf} \left[ \frac{1}{z_i-z_j} \right] \longrightarrow \text{Pf} \left[ \sum_{\mathbf{k}} g_{\mathbf{k}}\, \tilde\phi_{\mathbf{k}}\left(z_i\right)
\,\tilde\phi_{-\mathbf{k}}\left(z_j\right) \right],
\end{equation}
where $\tilde\phi_{\mathbf{k}}\left(z_i\right)= J_i^{-1}\mathcal{P}_\text{\tiny LLL} [\phi_{\mathbf{k}}\left(z_i\right)J_i ]$ are the projected CF orbitals, and $J_i = \prod_{k\neq i} \left( z_i - z_k\right)$.
On the sphere, the expansion in Eq.~(\ref{eq:generalG}) involves monopole harmonics (for details, see~\cite{Moller08b}, App.~A) and the number of relevant parameters $g_k$ is small (no more than five for the system sizes considered)~\cite{Moller08a}.

Fig.~\ref{fig:overlaps}b) shows the overlaps between the exact $\nu=12/5$ ground-state
at shift $S=2$ and the BS states with pair wave functions optimized to yield maximum overlap.
Comparing to the results in Fig.~\ref{fig:overlaps}a),
we find that the overlap peaks increase significantly, become wider, and shift to slightly higher $\delta V_1$. The overlaps now reach as high as $0.990(2)$ for $N_{e}=8$ and climb to $0.92(3)$ for our largest system ($N_{e}=14$) at $\delta V_1=0.02$.

In the region where the overlap with the BS state is non-zero, the spectrum has non-zero neutral gap [see Fig.~\ref{gaps}c)] above a homogeneous $L^{2}=0$ ground-state. The overlap of the BS state decreases for large $\delta V_1$, where the HH state is stabilized. In some cases (i.e. $N_{e}=10$ and $14$) the overlap drops discontinuously to zero around $\delta V_1 = 0.03$, indicating level crossings in the ground-state. This transition point slightly differs from that of the charge excitation gap shown in Fig.~\ref{gaps}b), which closes near $\delta V_1 = 0.01$. However, one may expect the $S=2$ charge gap to close prematurely as $\delta V_1$ approaches the region where the $S=4$ state is stabilized, because there will be competition at $S=3$ (which is used to evaluate the charge gaps) between the excited states of the BS and HH universality classes. Hence, this mismatch is a finite-size effect that should disappear in large systems where phase transitions are sharply defined.

The large overlaps between the gapped, incompressible FQH state that we found at $\nu=12/5$ with $S=2$ and the BS state's trial wave functions provide strong evidence of its universality class and establishes the BS state as a good candidate for the observed $\nu=12/5$ FQH state~\cite{Xia04,Pan08,Kumar10}, joining the ranks of $\overline{\text{RR}}$ and potentially HH as the primary contenders. Naturally, one would like to pit these states against each other in order to predict a favored candidate. However, directly comparing ground-state energies or gaps at a given value of $N_e$ small does not provide a valid assessment, as finite-size effects are significant. To be physically meaningful, energetics must be analyzed in the thermodynamic limit $N_e \rightarrow \infty$.

For very large $N_{e}$, the energetically favored universality class of a given filling fraction always wins out. In this case, changing $N_{\phi}$ results in the creation of quasiparticles, but the states remain in the same universality class. For the small values of $N_{e}$ considered in numerical studies, however, changing $N_{\phi}$ can have very significant effects. Indeed, we have demonstrated that the $\nu=12/5$ states with $S=2$ and $-2$ (a difference of $4$ fluxes) represent different universality classes for small system sizes.

Thus, we can compare the competing $\nu=12/5$ FQH states by extrapolating (from small $N_e$) the ground-state energy per electron of the Coulomb Hamiltonian for shifts $S=2$ and $-2$ to the thermodynamic limit. (We do not consider $S=4$ here, since it does not have a good FQH state at the Coulomb point.) To aid this finite size scaling analysis, we use the rescaled magnetic length $\ell_{0}^{\prime} = \sqrt{\nu \left( N_{\phi} -2 \right) / N_{e}} \ell_{0}$ (where the $-2$ accounts for being in the 2LL) and units of energy $e^2 / \ell_{0}^{\prime}$, which are meant to correct the effects of curvature in finite-sized spherical systems and make the energy behavior linear in $1/N_e$~\cite{Morf86}.
Using a least-squares fit to linearly extrapolate the rescaled ground-state energies per electron from $N_e=8,\ldots,16$ to the thermodynamic limit, we find $E/N_{e}= -0.3440(17)$ and $-0.3434(5)$ for $S=2$ and $-2$, respectively. These are very close and within extrapolation errors of each other, so a clear favorite cannot be identified. In principle, the thermodynamic extrapolation of ground-state energies could have given significantly different results for $S=2$ and $-2$, so we interpret this as an indication that the BS and $\overline{\text{RR}}$ states are in close competition.

In view of this, it is likely that physical effects not yet taken into account will play an important role in determining which state is actually favored and experimentally realized. In a first attempt to include some of these effects, we have modeled the system for finite well width $d$ by using pseudopotentials for an infinite square well in the spherical geometry. Results for the $N_{e}\rightarrow\infty$ extrapolation of the $\nu=12/5$ Coulomb ground-state energies with $S=2$ and $-2$ for well width $d=1,\ldots,3$ (measured in units of $\ell_{0}^{\prime}$) from system sizes $N_e=8,\ldots,16$ are given in Table~\ref{table-gse}. The $S=2$ and $S=-2$ extrapolated energies are always within extrapolation errors of each other, again demonstrating the close competition of the BS and $\overline{\text{RR}}$ states. Clearly, further numerical studies including Landau level mixing, spin, and a treatment of layer thickness that enlarges the Hilbert space with vertically excited subbands are desirable, but beyond the scope of this paper.

\begin{table}[t!]
\[
\begin{array}{|l|l|l|}
\hline
d        &   S=2               &   S=-2           \\
\hline
0        &   -0.3440(17)       &   -0.3434(5)       \\
1        &   -0.3100(14)       &   -0.3096(5)       \\
2        &   -0.2853(11)       &   -0.2852(5)       \\
3        &   -0.2657(9)        &   -0.2659(5)       \\
\hline
\end{array}
\]
\caption{Extrapolated ground-state energies for Coulomb interactions in finite wells of width $d$ at $\nu=12/5$ for $S=2$ and $-2$ (corresponding to the BS and $\overline{\text{RR}}$ states).}
\label{table-gse}
\end{table}

Another way to more directly compare different states of the same filling fraction is to examine them on the torus, where there is no shift. This was done in~\cite{Rezayi09a} for a particle-hole symmetric system at $\nu=13/5$ for $N_{e}=15$ and $18$. The results exhibit ground-state degeneracy on the torus that agrees with the $\overline{\text{HH}}$ state for most of the parameter space, and best agrees with the RR state in a small region near the Coulomb point. However, the gap is not large in this region, and close inspection reveals low lying states that may be identified as $\overline{\text{BS}}$ ground-states~\cite{Rezayi-private}. This again indicates that the inclusion of additional important physical effects may be significant in determining which state is actually energetically favored. Furthermore, no scaling analysis was carried out in~\cite{Rezayi09a}, so its prediction for the thermodynamic limit is unclear and a large $N_e$ level crossing in favor of another candidate is not ruled out.

It will be very interesting to see which state(s) experiments support as correctly describing the $\nu =12/5$ and other 2LL FQH plateaus. Given the results of our analysis, one may even speculate that more than one of the proposed states could turn out to be experimentally obtainable by realizing different physical regimes at $\nu=12/5$. Experiments that measure the electric charge of the fundamental quasihole will not distinguish between HH, BS, and $\overline{\text{RR}}$ for $\nu=12/5$, since these all have charge $e/5$ fundamental quasiholes. Experiments that probe scaling behavior or thermal conductance may potentially be able to distinguish between these states~\cite{Bishara08}, but are typically complicated by non-universal edge physics. Interference experiments, however, should be able to unambiguously distinguish between these possibilities~\cite{Bonderson06b,Bishara09}. Such interference experiments have recently been implemented for $\nu=5/2$~\cite{Willett09a}, providing evidence supporting a non-Abelian state. Hopefully, experimental evidence on the nature of $\nu=12/5$ will emerge soon.

\begin{acknowledgments}
We thank S.~Das~Sarma, C.~Nayak, S.~Simon, A.~W\'ojs, and especially E.~Rezayi for illuminating discussions. GM is supported by Trinity Hall Cambridge and I2CAM under NSF grant DMR-0844115. AF, GM, and JKS acknowledge the support and hospitality of Microsoft Station Q. JKS is supported by Science Foundation Ireland PI award 08/IN.1/I1961.
\end{acknowledgments}


\end{document}